\documentclass[reprint, prx, superscriptaddress, floatfix,longbibliography]{revtex4-2}
\usepackage{savesym}
\usepackage{hyperref}
\usepackage{graphicx}
\usepackage{amsmath}
\savesymbol{comment}
\usepackage{wasysym}
\usepackage{amsfonts}
\usepackage{color}
\usepackage{url}

\usepackage{verbatim}
\usepackage{upgreek}
\usepackage{bm}
\usepackage{units}
\usepackage{placeins}
\usepackage{babel}
\restoresymbol{TXF}{comment}
\usepackage{float}
\usepackage[draft]{changes}

\makeatother
\begin{document}
\title{Intrinsic energy flow in laser-excited 3{\sl d} ferromagnets}
\author{Daniela Zahn}
 \email{zahn@fhi-berlin.mpg.de}
\affiliation{Abteilung Physikalische Chemie, Fritz-Haber-Institut der Max-Planck-Gesellschaft, Faradayweg 4-6, 14195 Berlin, Germany}
\author{Florian Jakobs}
\affiliation{Dahlem Center for Complex Quantum Systems and Fachbereich
Physik, Freie Universit\"at Berlin, Arnimallee 14, 14195 Berlin, Germany}

\author{H\'{e}l\`{e}ne Seiler}
\affiliation{Abteilung Physikalische Chemie, Fritz-Haber-Institut der Max-Planck-Gesellschaft, Faradayweg 4-6, 14195 Berlin, Germany}

\author{\mbox{Tim A. Butcher}}
\altaffiliation[Present address: ]{Swiss Light Source, Paul Scherrer Institut, 5232 Villigen PSI, Switzerland}
\affiliation{Institute of Radiation Physics, Helmholtz-Zentrum Dresden-Rossendorf, Bautzner Landstra\ss{}e 400, 01328 Dresden, Germany}

\author{Dieter Engel}
\affiliation{Division B: Transient Electronic Structure and Nanophysics, Max-Born-Institut, Max-Born-Stra\ss{}e 2A, 12489 Berlin, Germany}

\author{Jan Vorberger}
\affiliation{Institute of Radiation Physics, Helmholtz-Zentrum Dresden-Rossendorf, Bautzner Landstra\ss{}e 400, 01328 Dresden, Germany}
\author{Unai Atxitia}
\affiliation{Dahlem Center for Complex Quantum Systems and Fachbereich
Physik, Freie Universit\"at Berlin, Arnimallee 14, 14195 Berlin, Germany}
\author{Yoav William Windsor}
\affiliation{Abteilung Physikalische Chemie, Fritz-Haber-Institut der Max-Planck-Gesellschaft, Faradayweg 4-6, 14195 Berlin, Germany}
\affiliation{Institut f\"ur Optik und Atomare Physik, Technische Universit\"at Berlin, Stra\ss{}e des 17.~Juni 135, 10623 Berlin, Germany}
\author{Ralph Ernstorfer}
 \email{ernstorfer@tu-berlin.de}
\affiliation{Abteilung Physikalische Chemie, Fritz-Haber-Institut der Max-Planck-Gesellschaft, Faradayweg 4-6, 14195 Berlin, Germany}
\affiliation{Institut f\"ur Optik und Atomare Physik, Technische Universit\"at Berlin, Stra\ss{}e des 17.~Juni 135, 10623 Berlin, Germany}

\begin{abstract}
Ultrafast magnetization dynamics are governed by energy flow between electronic, magnetic, and lattice degrees of freedom. A quantitative understanding of these dynamics must be based on a model that agrees with experimental results for all three subsystems. However, ultrafast dynamics of the lattice remain largely unexplored experimentally. Here we combine femtosecond electron diffraction experiments of the lattice dynamics with energy-conserving atomistic spin dynamics (ASD) simulations and {\sl ab initio} calculations to study the intrinsic energy flow in the 3{\sl d} ferromagnets cobalt (Co) and iron (Fe). The simulations yield a good description of experimental data, in particular an excellent description of our experimental results for the lattice dynamics. We find that the lattice dynamics are influenced significantly by the magnetization dynamics due to the energy cost of demagnetization. Our results highlight the role of the spin system as the dominant heat sink in the first hundreds of femtoseconds. Together with previous findings for nickel~[Zahn {\sl et al.}, \href{https://doi.org/10.1103/PhysRevResearch.3.023032}{Phys. Rev. Research 3, 023032 (2021)}], our work demonstrates that energy-conserving ASD simulations provide a general and consistent description of the laser-induced dynamics in all three elemental 3{\sl d} ferromagnets.
\end{abstract}

\maketitle
\date{\today}

\section{Introduction}
Ultrafast manipulation of magnetic order with light promises pathways to new applications in magnetic data storage and spintronics~\cite{2010Kirilyuk}. Femtosecond laser excitation can change magnetic order in various ways; for example, it can induce ultrafast demagnetization~\cite{1996Beaurepaire}, switch the magnetization direction~\cite{2011Radu,2021Stupakiewicz}, and induce spin reorientation~\cite{2021Afanasiev}. The microscopic mechanisms governing the response of magnetic materials to laser excitation continue to be a topic of current research~\cite{2010Koopmans,2010Battiato,2013Eschenlohr,Toews2015,2015Carpene,2017Eich,2018Tengdin,2018You,Dornes2019}. 
An important factor governing the response of a material to laser excitation is the intrinsic energy flow between electronic, magnetic, and lattice degrees of freedom. When Beaurepaire {\sl et al.} discovered ultrafast demagnetization in Ni, they introduced a phenomenological three-temperature model (3TM) to describe the observed magnetization dynamics \cite{1996Beaurepaire}. While the 3TM offers an intuitive explanation for the observed dynamics, recent studies suggest that it falls short of a full description of ultrafast demagnetization. 
In particular, there is experimental and theoretical evidence that the spin system is not in a thermal state on ultrafast timescales \cite{2018Tengdin,Kazantseva2007,2008Carpene}, suggesting that a more detailed description of the magnetic degrees of freedom is necessary.

To obtain a full quantitative description of a material's response to laser excitation, any proposed model must be verified by comparison to experimental data of the responses of electronic, magnetic, and lattice degrees of freedom. The lattice plays a major role in the dynamics of 3{\sl d} ferromagnets, since it drains energy from the electrons via electron-phonon coupling on similar timescales compared with the demagnetization, thus reducing the temperature of the electron system. On the other hand, lattice dynamics are also influenced by magnetization dynamics, even if the coupling is only indirect via the electron system. Our previous work on Ni demonstrated that energy flow into and out of the spin system leads to a significant slow-down of the lattice dynamics~\cite{our_nickel}. This suggests that accounting for this energy flow is integral to any model quantitatively describing the responses of all three subsystems in 3{\sl d} ferromagnets.

Despite their significant role in the energy flow dynamics, the lattice dynamics of 3{\sl d} ferromagnets are less studied compared with electron and spin dynamics~\cite{2008Carpene,2010Koopmans,2015Carpene,2017Eich,2018Gort,2018Buehlmann}. Time-resolved diffraction offers the most direct way to study lattice dynamics since it is only sensitive to the lattice. Hitherto, only two studies of the sub-picosecond lattice dynamics of Co or Fe with time-resolved diffraction exist~\cite{Durham2020,Rothenbach2019} and neither of them focuses on the lattice heating in the ferromagnet. Furthermore, literature values for the electron-phonon coupling parameter $G_\mathrm{ep}$ vary significantly, from $6\times10^{17}$ to $\unit[4.05\times10^{18}]{\frac{W}{m^3K}}$ for Co \cite{2005Bigot,2010Koopmans,2019Ritzmann,2021Unikandanunni} and from $7\times10^{17}$ to $\unit[5.48\times10^{18}]{\frac{W}{m^3K}}$ for Fe \cite{Lin_website,2019Ritzmann,2020Medvedev,2017Migdal,2013Petrov,2018Ogitsu}. In addition, there are several literature values for the electron-phonon coupling parameter $\lambda$, which is related to $G_\mathrm{ep}$ (see, for example, Ref.~\onlinecite{2008Lin}) and also varies significantly \cite{Papaconstantopoulos_handbook,Allen_lambdas,1984Jarlborg,2013Verstraete}. In ferromagnets, extracting the electron-phonon coupling solely from experiments is particularly challenging because three different subsystems contribute to the observed dynamics.

Here we measure the lattice dynamics of Co and Fe directly using femtosecond electron diffraction. Instead of extracting $G_\mathrm{ep}$ from experiments, we perform spin-resolved density functional theory (DFT) calculations, which yield $G_\mathrm{ep}$ as well as the heat capacities of the electrons and the lattice \cite{2016Waldecker,our_nickel}.
Based on the experimentally measured lattice dynamics and the DFT results, we study the intrinsic energy flow between electronic, magnetic, and lattice degrees of freedom. We employ energy-conserving atomistic spin dynamics (ASD) simulations \cite{Wienholdt2015,our_nickel}, a hybrid model which combines conventional ASD simulations with a description of the energy flow between all subsystems. By directly simulating the evolution of the spin system, ASD simulations have the advantage that they are not constrained to thermal descriptions of the spin system. Previously, we applied this approach to Ni with excellent agreement between theory and experiment \cite{our_nickel}. Here we demonstrate that the same considerations hold also for Co and Fe, thus generalizing our approach to all three elemental 3{\sl d} ferromagnets. To demonstrate the strong influence of the magnetization dynamics on the lattice dynamics, we compare results of the conventional two-temperature model (TTM), which does not consider the spin system, to results of the energy-conserving ASD simulations. With the latter, we obtain excellent agreement with the lattice dynamics of Co and Fe as well as a good description of the magnetization dynamics. This demonstrates that ASD simulations offer a consistent description of the laser-induced dynamics in all three elemental 3{\sl d} ferromagnets.

In Section~\ref{sec:experiment}, we describe the experiment and the data analysis. Section~\ref{sec:results} presents both experimental results for the lattice dynamics and model results. Based on the ASD simulation results, in Section~\ref{sec:discussion} we discuss the intrinsic energy flow between electrons, spins, and the lattice in detail. Section~\ref{sec:summary} summarizes the main findings. 

\section{Time-resolved diffraction experiment}
\label{sec:experiment}
 The samples of our experiments are freestanding thin films of Co or Fe with a thickness of \unit[20]{nm}, sandwiched between \unit[5]{nm}-thick layers of silicon nitride. They were grown on a single crystal of NaCl by magnetron sputtering, resulting in polycrystalline films. Next, they were transferred onto a transmission electron microscopy (TEM) grid by floating the films on water. The samples were not subjected to a magnetic field before the experiment, i.e., different magnetic domains are likely present in the sample.

\begin{center}
\begin{figure*}[bth]

\includegraphics[width=\linewidth]{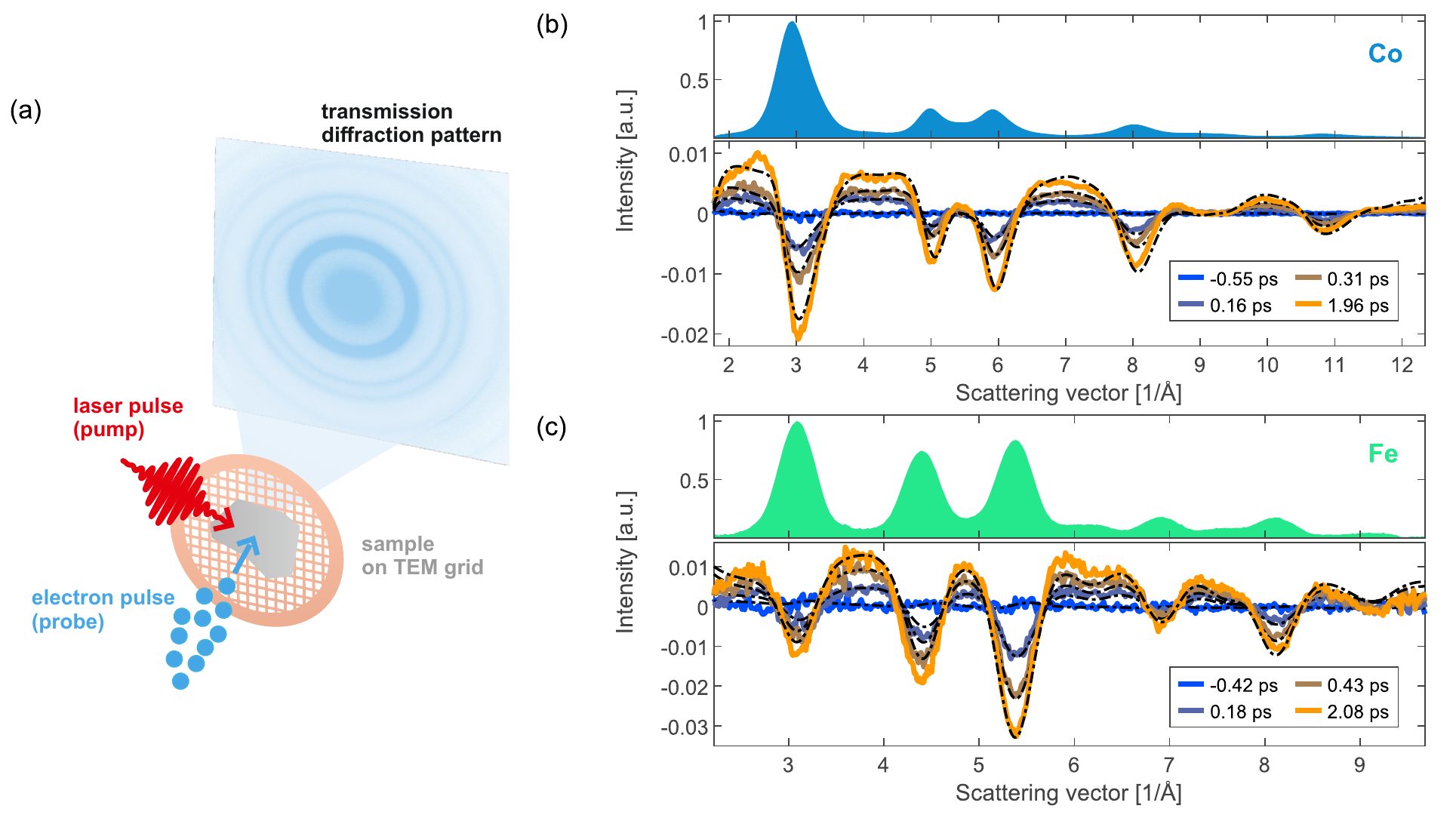}
\caption{The femtosecond electron diffraction experiment. (a) Schematic illustration of the measurement. The samples are thin, freestanding films on TEM grids, which are excited by ultrashort laser pulses. The lattice response is probed using ultrashort electron pulses, which diffract off the sample. Diffraction patterns are recorded in transmission. In the case of polycrystalline samples such as the samples studied in this work, the diffraction patterns consist of rings. A diffraction pattern of our Co sample is shown. (b) Diffraction pattern of Co and time-resolved changes. The upper part shows the azimuthally averaged diffraction pattern (radial profile, RP) of Co. Here the background-subtracted pattern is shown for illustrational purposes; however, note that in the analysis of the diffraction patterns, fits are performed to background and rings simultaneously. The lower part shows the differences of the RPs compared with the RP before laser excitation for several pump-probe delays (solid curves). The dashed black lines show the fit results of the global fitting routine described in Sec.~\ref{sec:experiment} and in detail elsewhere \cite{our_Pt}. (c) Same as (b) but for Fe.}
\label{fig1:intro}

\end{figure*}
\end{center}
To directly access the lattice dynamics after laser excitation, we employ femtosecond electron diffraction using the setup described in Ref.~\onlinecite{2015Waldecker}. A schematic illustration of the experiment is presented in Fig.~\ref{fig1:intro}(a). In the electron diffraction experiment, the thin films are excited with an ultrashort laser pulse. The lattice response to laser excitation is probed using an ultrashort high-energy electron pulse. The electrons diffract off the sample and are recorded in transmission. The electron energy was \unit[70]{keV} for the experiments on Co and \unit[60]{keV} for the experiments on Fe. All experiments were performed at room temperature (\unit[295]{K}). Since the samples are polycrystalline, the diffraction patterns consist of Debye-Scherrer rings, as shown exemplarily in Fig.~\ref{fig1:intro}(a) for our Co sample. Our main observables are changes in the intensities of the diffraction rings following laser excitation. These are directly related to the change in atomic mean-squared displacement (MSD)~\cite{Peng}:
\begin{equation}
    \frac{I(t)}{I_0}=\mathrm{exp}\{ -\frac{1}{3}\hspace{2pt}q^2\hspace{2pt}\Delta \langle u^2\rangle\ \}.
\end{equation}
Here, $q$  is the scattering vector of the diffraction ring ($q=4\pi\hspace{1pt}\mathrm{sin}(\theta)/\lambda$), $\Delta \langle u^2\rangle=\langle u^2\rangle(t)-\langle u^2\rangle(t<0)$ is the MSD change, $I(t)$ is the intensity as a function of pump-probe delay, and $I_0$ is the intensity before laser excitation.

To extract the MSD dynamics from the diffraction patterns, we employ a global two-step fitting routine \cite{our_Pt}. In brief, the first step is a fit to the diffraction pattern before laser excitation. The fit function consists of a background function plus Lorentzians for the diffraction rings, all convolved with a Gaussian to account for the finite coherence of the electron beam. In the second step, the time-dependent changes are extracted. For this, we fix most parameters of the fit function and allow only changes of the lattice constant (i.e., expansion/contraction of the lattice), changes of the MSD, and changes of the background parameters. The lattice dynamics are extracted from the full diffraction pattern instead of individual diffraction rings, which increases the reliability of the results. Further information on the global fitting routine is available in Ref.~\onlinecite{our_Pt}. 

\section{Results}
\label{sec:results}
\subsection{Experimental results for the lattice dynamics}
\begin{figure}[bth]
\begin{center}
\includegraphics[width=\columnwidth]{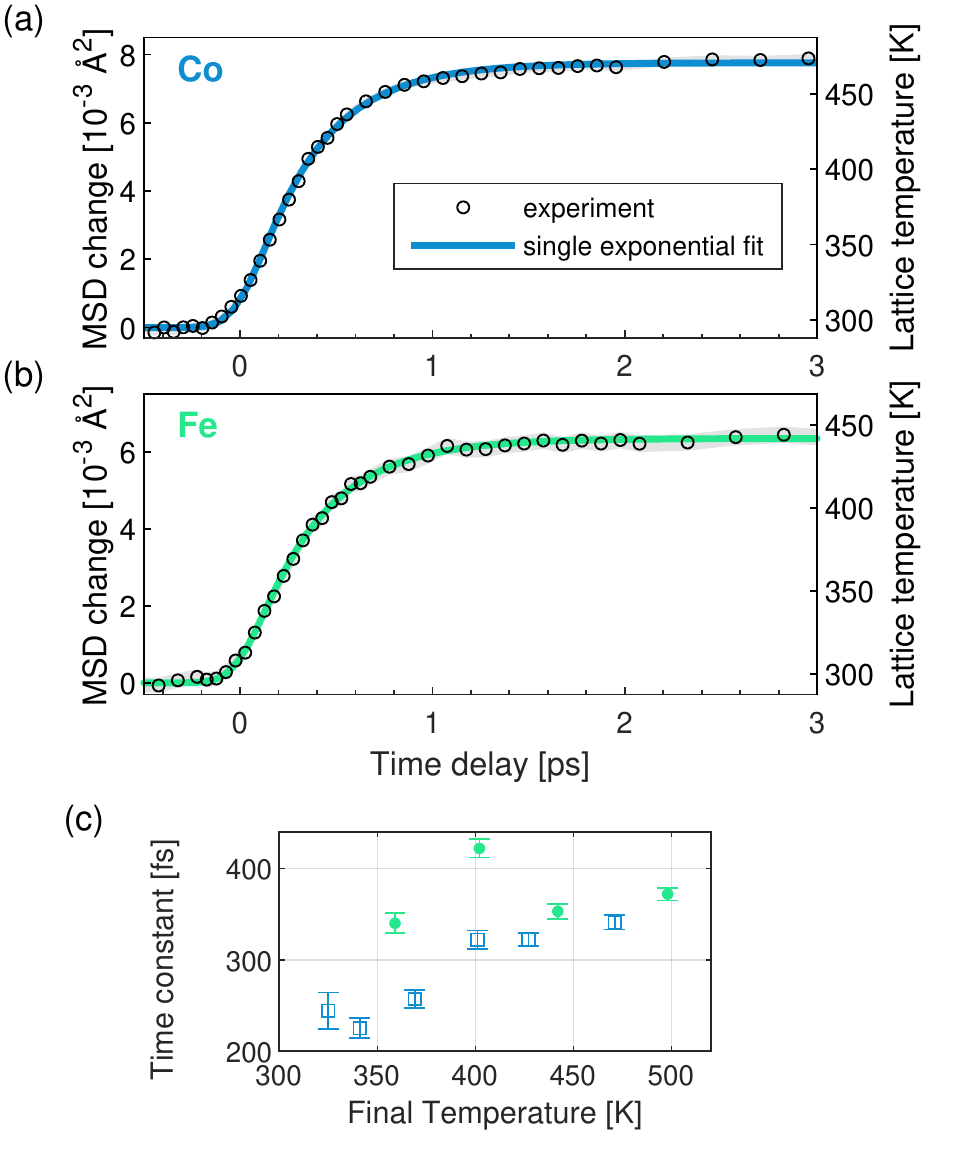}
\caption{Experimental results for the lattice dynamics and single exponential fits. (a) Evolution of the atomic MSD and corresponding lattice temperature for Co. (b) MSD evolution and corresponding lattice temperature for Fe. The solid lines in (a) and (b) are the results of fits to the experimental data with a single exponential function, convolved with a Gaussian (\unit[250]{fs} FWHM) to account for the time resolution of the experiment. The excitation wavelength was \unit[2300]{nm}. (c) Fit results for the time constant of the single exponential function for different excitation densities, yielding different final lattice temperatures.}
\label{fig2:exp_results}
\end{center}
\end{figure}

Experiments were performed on Co and Fe for several excitation densities each. For every excitation density, several delay scans were recorded and the results were averaged before applying the two-step fitting routine. Examples for the resulting MSD dynamics of Co and Fe are presented in Figs.~\ref{fig2:exp_results}(a) and \ref{fig2:exp_results}(b), respectively. For the conversion of MSD to lattice temperature, we calculated the temperature-dependent Debye-Waller factors for Fe and Co based on the phonon density of states (DOS) from DFT (see Appendix~A). 
We performed fits to the experimental data using a single exponential function, convolved with a Gaussian of \unit[250]{fs} (FWHM) to account for the time resolution. The time constant of the single exponential function, the amplitude, and the onset (time zero) were fit parameters and the fit range was from -0.5 to \unit[4]{ps}. 
The results for the time constant are shown in Fig.~\ref{fig2:exp_results}(c) for different excitation densities. For Co, we find that the time constant increases with increasing excitation density. For Fe, no clear trend is observed. 

\subsection{Comparison of the experimental results to energy flow models}
\subsubsection{TTM}
\begin{figure}[bth]
\begin{center}
\includegraphics[width=\columnwidth]{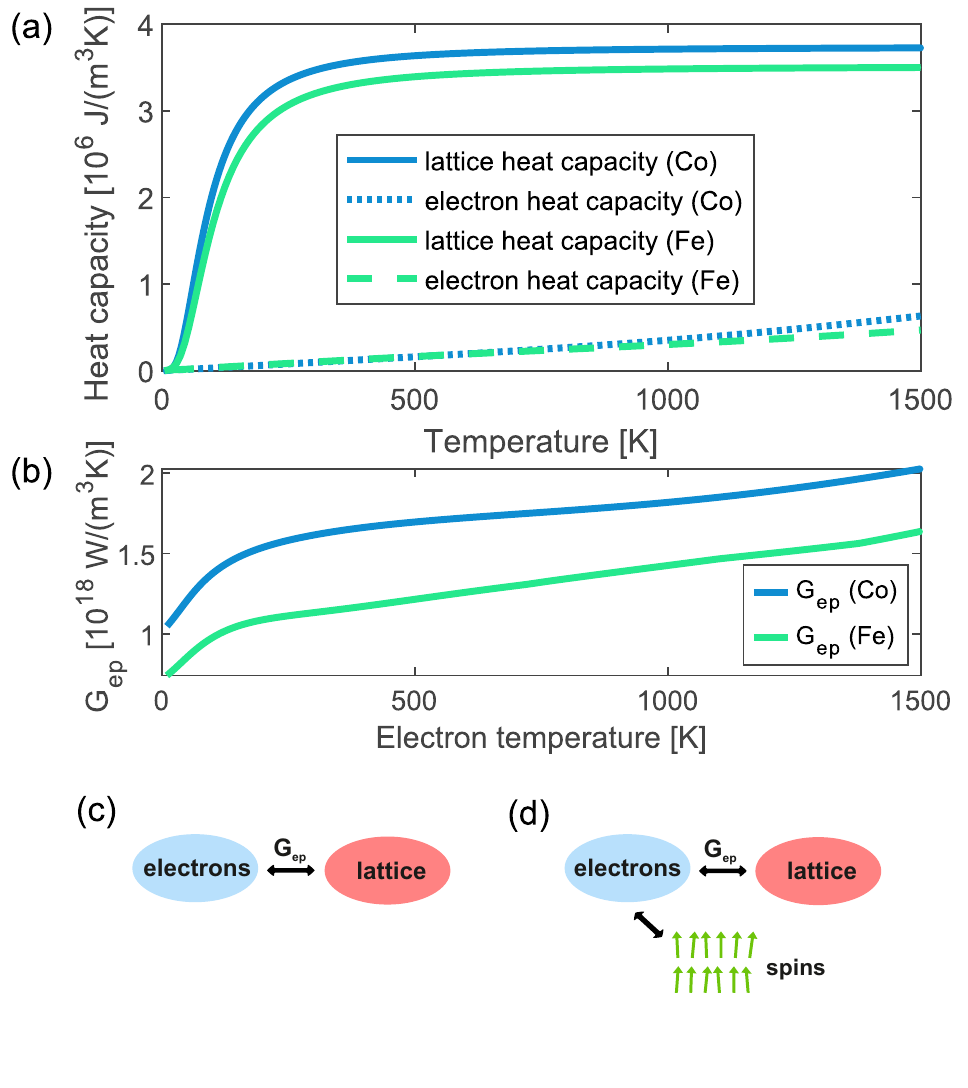}
\caption{Heat capacities, electron-phonon coupling parameters, and schematic illustrations of the employed energy flow models. (a) Electronic (dashed curves) and lattice (solid curves) heat capacities, and (b) electron-phonon coupling parameters $G_\mathrm{ep}$ as a function of electron temperature calculated from spin-resolved DFT results.}
\label{fig3:heat_capacities}
\end{center}
\end{figure}

In the next step, our goal is to analyze the intrinsic energy flow between electronic, magnetic, and lattice degrees of freedom. For this, we compare our experimental data to models for the energy flow. In order to minimize the number of free parameters in the models, we use spin-resolved DFT to obtain the (electron-temperature-dependent) electron-phonon coupling parameter as well as the electron and lattice heat capacities. The results for the heat capacities and the electron-phonon coupling parameters are presented in Figs.~\ref{fig3:heat_capacities}(a) and \ref{fig3:heat_capacities}(b). All electronic heat capacity and $G_\mathrm{ep}$ curves presented in Fig.~\ref{fig3:heat_capacities} are the sum of majority and minority carrier contributions. Details about the DFT calculations are described in Appendix~A.

\begin{center}
\begin{figure*}[bth]

\includegraphics[width=\linewidth]{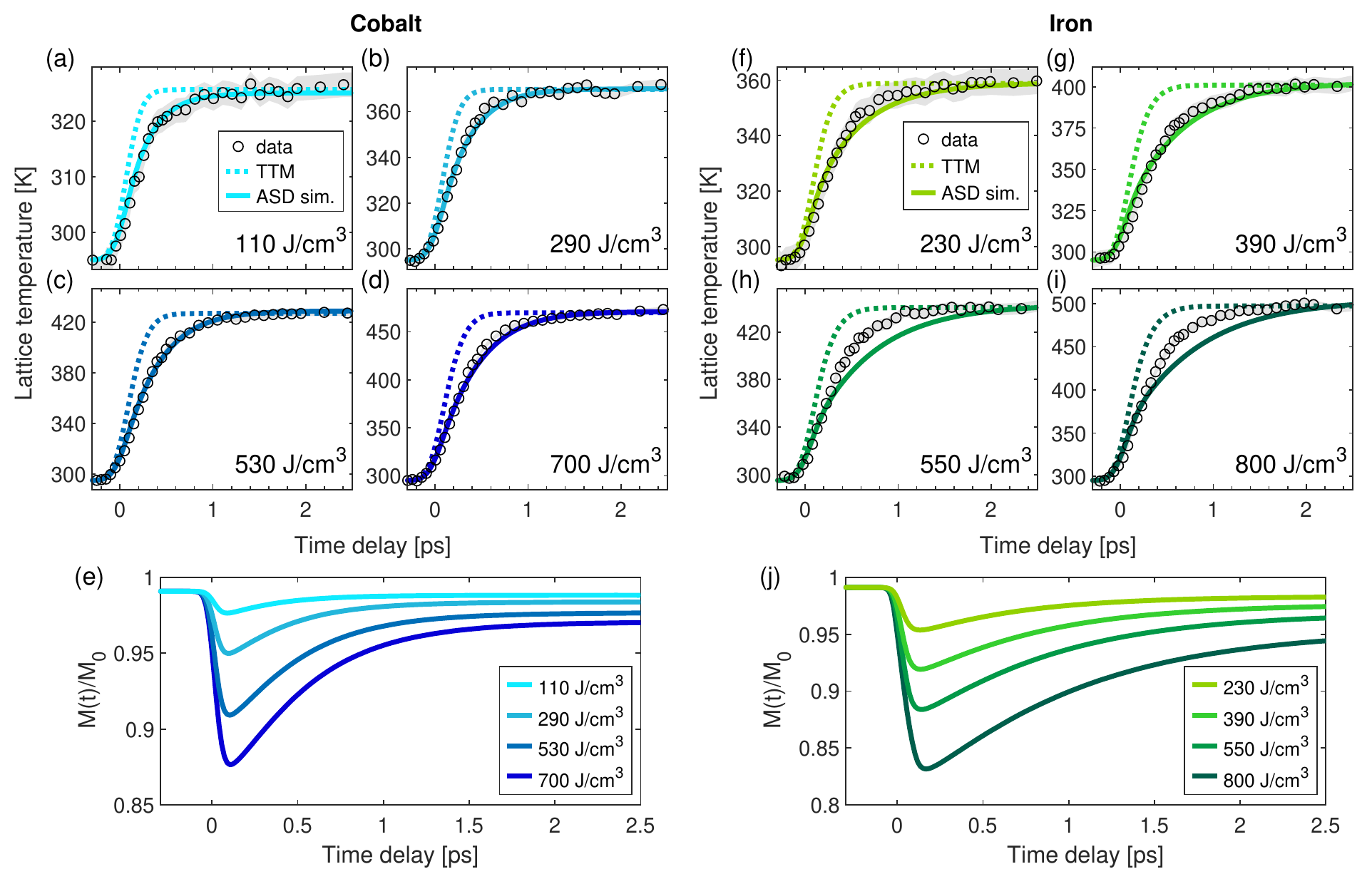}
\caption{Experimentally measured lattice dynamics and model predictions. (a)-(d) Lattice temperature as a function of pump-probe delay in Co for different absorbed energy densities. The experimental data are shown as black circles. The results of the TTM are shown as dashed curves and the results of the ASD simulations are shown as solid curves. The TTM and ASD results were convolved with a Gaussian with a FWHM of $\unit[\sqrt{250^2-80^2}]{fs}\approx\unit[237]{fs}$, which accounts for the temporal broadening induced by the probe pulse (the effect of the pump pulse width is already included in the model itself). The gray-shaded areas represent the standard errors of the experimental data, obtained from the fitting routine described in Section~\ref{sec:experiment}. The displayed energy densities correspond to the absorbed energy densities in the ASD simulations. (e) Magnetization dynamics of Co predicted by the ASD simulations. (f)-(i) Experimental results for the lattice temperature in Fe alongside results of the TTM and the ASD simulations. (j) Magnetization dynamics of Fe predicted by the ASD simulations.}
\label{fig4:TTM_ASD}

\end{figure*}
\end{center}

Having obtained the input parameters for the models from DFT, we start with the conventional TTM \cite{1974Anisimov,1987Allen}, which considers only the electronic and lattice degrees of freedom but disregards the spin system. The system is modeled as two heat baths, electrons and the lattice, which are coupled by the electron-phonon coupling parameter $G_\mathrm{ep}$. The evolution of the electron temperature ($T_\mathrm{e}$) and the lattice temperature ($T_\mathrm{l}$) is then described by two coupled differential equations:
\begin{equation}
c_\mathrm{l}(T_\mathrm{l})\cdot\frac{dT_\mathrm{l}}{dt}=G_\mathrm{ep}(T_\mathrm{e})\left[T_\mathrm{e}-T_\mathrm{l}\right]
\label{eq:TTM1}
\end{equation}
\begin{equation}
c_\mathrm{e}(T_\mathrm{e})\cdot\frac{dT_\mathrm{e}}{dt}=G_\mathrm{ep}(T_\mathrm{e})\left[T_\mathrm{l}-T_\mathrm{e}\right]+P(t).
\label{eq:TTM2}
\end{equation}
Here $c_\mathrm{e}$ and $c_\mathrm{l}$ are the electronic and lattice heat capacities, and $P(t)$ is the source term, i.e., the energy input to the electronic system due to the laser excitation. The laser excitation is modeled as a Gaussian with a FWHM of \unit[80]{fs}. Its maximum (time zero) is determined from the single exponential fits described earlier. The energy deposited by the laser is determined from the lattice temperature after electron-lattice equilibration (in the range from \unit[1.5 to 4]{ps} after laser excitation) and the heat capacity (sum of electron and lattice contribution). Hence, there are no fit parameters in this TTM. The comparison between the TTM and the experimental results for the lattice dynamics is shown in Fig.~\ref{fig4:TTM_ASD} for both materials and several fluences each (dashed curves). We find that for both Fe and Co, the lattice temperature rise predicted by the TTM is faster compared with our experimental results. This finding agrees with previous results on Ni \cite{our_nickel}. A major source of this disagreement is the fact that the TTM does not consider magnetic degrees of freedom. Therefore, also the energy associated with magnetization dynamics is neglected. However, as we showed previously for the case of Ni, energy flow into and out of magnetic degrees of freedom has a profound influence on lattice dynamics~\cite{our_nickel}. Hence a model that takes the spin system into account is needed.

\subsubsection{ASD simulations}
In order to include the spin system in our model of the energy flow dynamics, we use energy-conserving ASD simulations, which simulate the dynamics of the spin system based on a Heisenberg model and the stochastic Landau-Lifshitz-Gilbert (s-LLG) equation. The coupling of electron and phonon system is described with a TTM based on the DFT results [see Figs.~\ref{fig3:heat_capacities}(a) and \ref{fig3:heat_capacities}(b)], as in the previous subsection. Energy conservation is achieved by monitoring the energy content of the spin system and subtracting/adding the change in spin energy from/to the electron system at each time step of the simulation. The TTM equation for the electron temperature (Eq.~\ref{eq:TTM2}) is thus modified as follows:
\begin{equation}
c_\mathrm{e}\hspace{1pt}\frac{\Delta T_\mathrm{e}}{\Delta t}=G_\mathrm{ep}\hspace{1pt}(T_\mathrm{l}-T_\mathrm{e})+P(t)-\frac{\Delta E_\mathrm{s}}{\Delta t}.
\label{eq:TTM_ASD}
\end{equation}
Here $\Delta E_\mathrm{s}$ corresponds to the change of spin energy in the time step ${\Delta t}$. It is calculated as follows:
\begin{equation}
\Delta E_\mathrm{s} = \frac{s^2}{s(s+1)}(\mathcal{H}\{\mathbf{S}_i(t + \Delta t)\} - \mathcal{H}\{\mathbf{S}_i(t)\}).
\label{eq:spin_energy}
\end{equation}
Here the $\mathbf{S}_i$ are the individual spins of the ASD simulation and the factor $s^2/\left[s(s+1)\right]$ accounts for the quantized nature of the spins ($s\approx \frac{3}{2}$ for Co and $s\approx 2$ for Fe). Note that direct spin-lattice coupling is not included in the model. The fast demagnetization timescales in 3{\sl d} ferromagnets suggest that the magnetization dynamics are dominated by electron-spin coupling. In nickel, spin-lattice coupling was estimated to be an order of magnitude smaller than other coupling constants~\cite{1996Beaurepaire}. More details about the energy-conserving ASD simulations are described in Ref.~\onlinecite{our_nickel} and the material-specific simulation parameters for Co and Fe are stated in Appendix~B. With this model, both the nonequilibrium spin dynamics and the energy flow between electrons, spins, and the lattice can be described.

The coupling between electrons and spins in the ASD simulations is governed by the damping parameter $\alpha$. It determines how fast the spins react to the stochastic field of the s-LLG equation, whose amplitude in turn depends on the electronic temperature. 
Here, we use $\alpha=0.01$ for Co and $\alpha=0.005$ for Fe, which yield a good description of the experimentally measured lattice dynamics at low excitation densities as well as realistic magnetization dynamics. These values are in good agreement with recent experimental results for $\alpha$~\cite{2021Mohan}. 

Figure~\ref{fig4:TTM_ASD} presents the ASD simulation results for both Co and Fe. First, we focus on the results for Co, shown in Figs.~\ref{fig4:TTM_ASD}(a)--\ref{fig4:TTM_ASD}(e). We find excellent agreement with the experimentally measured lattice dynamics for all excitation densities. Clearly, the agreement is much better than that obtained with the TTM. This finding highlights the importance of considering energy flow into and out of magnetic degrees of freedom, in agreement with our previous results for Ni \cite{our_nickel}.

The ASD simulation results for the magnetization dynamics of Co are presented in Fig.~\ref{fig4:TTM_ASD}(e). The general shape of the magnetization dynamics, in particular the pronounced drop and relatively fast recovery of the magnetization, agrees well with recent experimental results \cite{2021Unikandanunni,2020Borchert}. Regarding the demagnetization dynamics in the first hundreds of femtoseconds, the ASD simulation results reach the minimal magnetization roughly \unit[100--200]{fs} faster than in reported experiments~\cite{2021Unikandanunni,2020Borchert,2010Koopmans}. 
This could be due to deviations of the electronic distribution from a Fermi-Dirac distribution at early times after laser excitation, and due to the phenomenological electron-spin coupling in the ASD simulations. In addition, the ASD simulations describe an idealized system without defects or surface effects and assume homogeneous excitation, which can also contribute to the observed discrepancies. Regarding the magnetization recovery, we observe good agreement with results from Ref.~\onlinecite{2021Unikandanunni} while the recovery measured by Refs.~\onlinecite{2020Borchert,2010Koopmans} is slower than the ASD simulation results.

It should be noted that there is some spread in the experimental results for the magnetization dynamics, even when only thin films on non-metallic substrates are considered~\cite{2020Borchert,2010Koopmans,2021Unikandanunni,2009Krauss,2016Turgut,2021Mohan}. On short timescales, the measured results can contain artifacts from state-filling effects when probing optically~\cite{2000Koopmans,2017Razdolski}. On longer timescales, magnetization dynamics can be influenced by transport effects (of electrons and phonons out of the probed region), which depend on the sample geometry. Also other macroscopic sample properties may play a role in the magnetization response. A recent study found differences in the ultrafast response depending on the orientation of the magnetization relative to the crystal lattice~\cite{2021Unikandanunni}. In principle, both the demagnetization and the magnetization recovery contain valuable information on the coupling strength between electrons and spins. For example, reducing $\alpha$ in the ASD simulations leads to a slower demagnetization but also to a less pronounced magnetization recovery because the spin system heats less (and thus absorbs less energy) during the time when the lattice is still cold. A more precise comparison of model results to the responses of all subsystems could be obtained by measuring the lattice, magnetization, and electron dynamics on identical samples.

Next, we focus on the ASD simulation results for Fe, shown in Figs.~\ref{fig4:TTM_ASD}(f)--\ref{fig4:TTM_ASD}(j). For low fluences, we obtain excellent agreement with the experimentally measured lattice dynamics, again corroborating the strong influence of the spin dynamics on the lattice dynamics. However, the quality of agreement is not as high as for Co. Specifically, for high fluences, the simulations predict lattice dynamics that are slower than the experimental observations.

In the following, we discuss possible reasons for these deviations. In our ASD simulations, the strength of the electron-spin coupling, described by the damping parameter $\alpha$, is constant. At higher excitation densities, however, the electron-spin coupling could react to the laser-induced changes of the electronic structure. Since Fe has the largest spin heat capacity of all three elemental 3{\sl d} ferromagnets at room temperature in combination with a rather low electronic heat capacity, its lattice dynamics are most sensitive to energy flow into and out of the spin system. Therefore, it is plausible that deviations between ASD simulations and experiments performed at high fluences are larger for Fe compared with Ni or Co. Furthermore, transient nonthermal electron and phonon distributions could contribute to the observed lattice dynamics for both Fe and Co \cite{2019Ritzmann,2020Wilson}. Experimentally, we observed a small apparent shift in time zero by tens of fs for high excitation densities. This could be caused by electron thermalization, which is more efficient at high excitation densities and typically enhances energy transfer to the lattice~\cite{2013Mueller}. Nonthermal distributions of electrons and phonons are not accounted for by our models and including them might change the optimal $\alpha$ toward lower values. In addition, direct spin-lattice coupling is not included in our model, as mentioned above. Even though we expect this coupling to be weak, it constitutes another channel for energy flow to the lattice and could enhance in particular the energy flow out of the spin system. Finally, DFT calculations are ground-state calculations. After laser excitation, band structure changes (for example a transient reduction of the exchange splitting) can occur~\cite{2018Tengdin}, which lead to changes of the electronic heat capacity and the electron-phonon coupling, especially for higher fluences. Hence, ASD simulations are expected to be most accurate for low excitation densities in general, which we observed also for Ni~\cite{our_nickel}. Nevertheless, for low and moderate fluences, our ASD simulations offer an excellent description of the laser-induced lattice dynamics for all three 3{\sl d} ferromagnets.

Regarding the magnetization dynamics of Fe, the ASD simulation results are presented in Fig.~\ref{fig4:TTM_ASD}(j). The initial demagnetization rate agrees well with experimental results~\cite{2020Borchert}. For the magnetization recovery, different results are reported in the literature~\cite{2020Borchert,2020Zhang,2008Carpene,2015Carpene,2017Razdolski,2018Buehlmann,2021Mohan}, from very little or no recovery~\cite{2020Borchert,2017Razdolski} to almost complete recovery~\cite{2020Zhang} on few-picosecond timescales. Only thin films on non-metallic substrates are considered here, which are expected to have the least transport effects. Due to the large spread of literature results, as in the case of Co, a more precise comparison of the model to the results of all three subsystems would require measuring their responses on identical samples. Based on the available experimental data, we conclude that our simulations provide a realistic description of the magnetization dynamics. Energy-conserving ASD simulations thus offer a description that is consistent with the responses of the lattice and the magnetization, which is an important step toward a complete, consistent description of the laser-induced dynamics of 3{\sl d} ferromagnets.

\section{Discussion}
\label{sec:discussion}
\begin{figure}[t!]
\begin{center}
\includegraphics[width=\columnwidth]{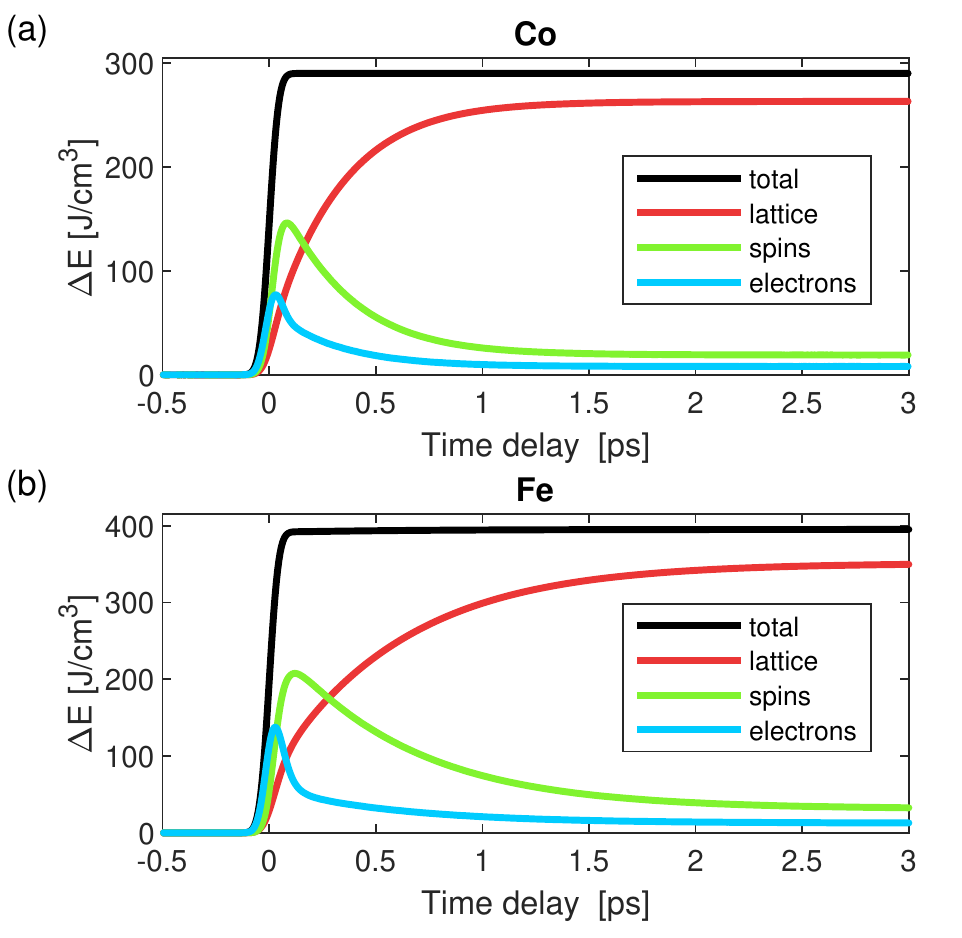}
\caption{Intrinsic energy flow between electrons, spins, and the lattice for (a) Co and (b) Fe. Here results with the same excitation density as in Fig.~\ref{fig4:TTM_ASD}(b) (Co) and Fig.~\ref{fig4:TTM_ASD}(g) (Fe) are presented. The additional energy density $\Delta E$ in the system is displayed. After laser excitation, the total energy (black) stays constant and energy is redistributed between electronic (blue), magnetic (green), and lattice (red) degrees of freedom.}
\label{fig5:energy}
\end{center}
\end{figure}

The good agreement of the ASD simulations with our experiments and the disagreement of the TTM show that energy flow into and out of the spin system has a significant impact on the lattice dynamics of Co and Fe. Based on the ASD simulation results, we are now able to analyze the intrinsic energy flow between electronic, magnetic, and lattice degrees of freedom in detail. The distribution of the absorbed energy between the three subsystems is presented in Fig.~\ref{fig5:energy}. 

After laser excitation, the total energy in the system stays constant, which is visualized by the black curve. From then on, energy is only redistributed between the different degrees of freedom. The laser pulse excites the electrons (blue curve), which initiates the energy flow from the electrons to the spin system (green curve). Already shortly after excitation, the spin system contains more of the additional energy than the electron system. Once spins and electrons have equilibrated and the electrons cool down further due to electron-phonon coupling, energy starts flowing back from the spin system to the electrons. In addition, energy also flows from the electrons to the lattice, such that in total, energy flows out of the electron system, although at a lower rate than during the demagnetization. Finally, thermal equilibrium is established after several picoseconds.

\begin{figure}[t!]
\begin{center}
\includegraphics[width=\columnwidth]{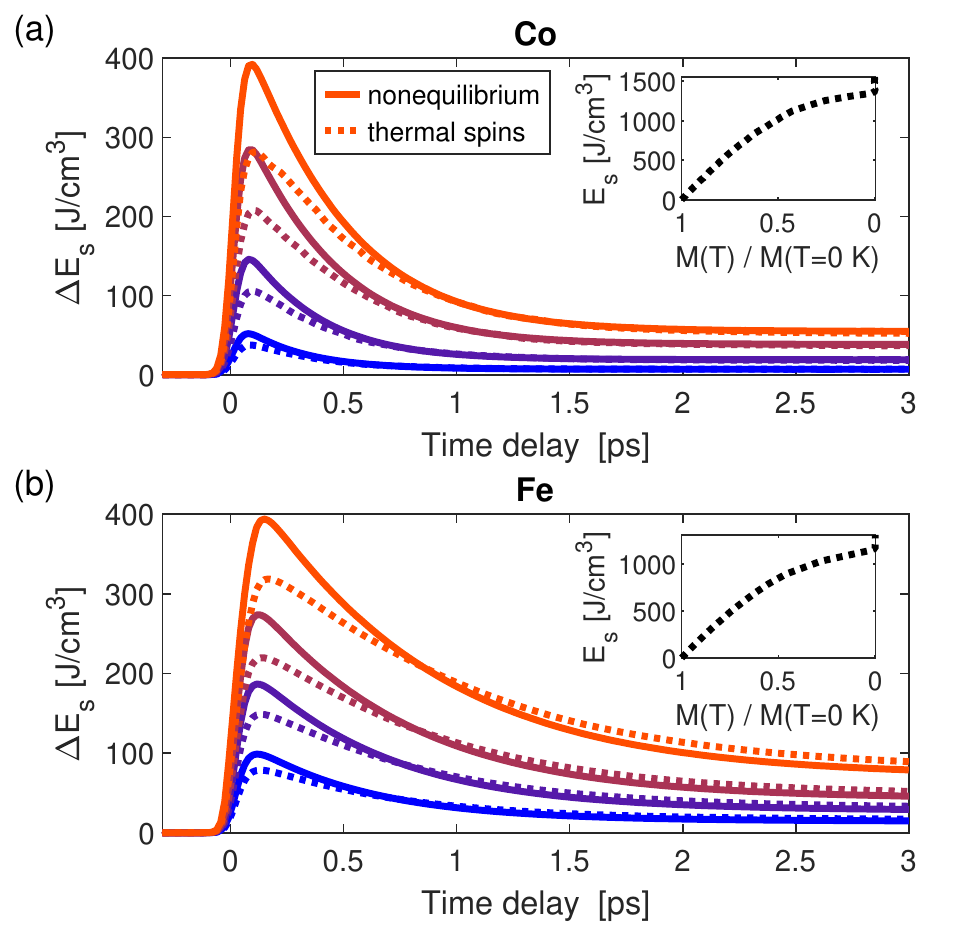}
\caption{Nonthermal spin dynamics for (a) Co and (b) Fe. The solid curves show the additional energy content of the spin system $\Delta E_\mathrm{s}$ as a function of pump-probe delay for the same fluences as in Fig.~\ref{fig4:TTM_ASD}. In contrast, the dashed curves show the additional spin energy content of a hypothetical, thermalized spin system with the magnetization dynamics from the ASD simulations, which was calculated using the equilibrium relationship between spin energy and magnetization. Differences between the solid and the dashed curves indicate a nonthermal spin system. Note that the very small differences that persist on timescales larger than \unit[3]{ps} are numerical artifacts. They could stem from the finite time steps in the nonequilibrium simulations or the larger $\alpha$ employed in the simulations of quasistatic heating. The insets show the equilibrium relationships between spin energy and magnetization.}
\label{fig6:nonthermal}
\end{center}
\end{figure}

Similar to our previous results for Ni, we find that also for Fe and Co, the ASD simulations predict a nonthermal spin system on short timescales after laser excitation. This is presented in Fig.~\ref{fig6:nonthermal}. The additional spin energy in the system is shown as solid curves (the absorbed energy densities are the same as in Fig.~\ref{fig4:TTM_ASD}). In addition, the dashed curves show how a thermalized spin system would behave. The thermalized case is based on the equilibrium properties of the spin system and the magnetization dynamics from the nonequilibrium simulation. We use the equilibrium relationships between magnetization and spin energy, shown in the insets of Fig.~\ref{fig6:nonthermal}, to translate the magnetization dynamics from the simulations into spin energy dynamics. Comparing this result to the spin energy dynamics obtained directly from the simulations allows to identify deviations from thermal behavior: Whenever the two quantities do not coincide, the spin system is in a nonthermal state. 

On short timescales below ca.~\unit[1]{ps}, dashed and solid curves differ, which indicates that the spin system is in a nonthermal state during this period. This nonthermal state is characterized by a relatively high spin energy content compared with the demagnetization amplitude, as the comparison between dashed and solid curves directly shows. This is analogous to our ASD simulation results for Ni and indicates that relatively many spin excitations with significant misalignment of neighboring spins are present compared with thermal equilibrium, which cost a lot of energy per magnetization reduction. The finding is corroborated by inspecting the simulated spin configuration at short time delays (see Appendix~B), which exhibits disorder on small length scales, i.e. also between spins that are close to each other. For the fluences reached in our experiments, the spin system thermalizes within the first picosecond after laser excitation. In contrast, for Ni, we observed a prolonged nonthermal behavior for higher fluences. These differences between  Fe, Co, and Ni are caused by their different Curie temperatures. Ni has a Curie temperature of only \unit[631]{K}, while the Curie temperatures of Fe and Co are \unit[1044]{K} and \unit[1390]{K}, respectively \cite{1971Crangle}. As a consequence, for the same absorbed energy density, Ni demagnetizes much more than Fe or Co \cite{2020Borchert}. For stronger demagnetization of Fe or Co, a prolonged nonthermal behavior is observed as well, which is shown exemplarily for Fe in Appendix~B. The prolonged nonthermal behavior is found to be caused by domain formation during the remagnetization process, in agreement with previous results by Kazantseva {\sl et al.}~\cite{Kazantseva2007}. The ASD simulation results thus suggest that in particular for strong demagnetization, a thermal description of the spin system is not adequate to describe nonequilibrium dynamics of ferromagnets.

In summary, according to the ASD simulations, two key aspects of ultrafast energy flow dynamics in ferromagnets are apparent: the slowdown of the lattice response caused by energy flow into and out of the spin system, and the nonthermal behavior of the spin system on short timescales. The first effect can be qualitatively reproduced with a simple 3TM as well (depending on the coupling constants, which are typically fit parameters). However, the second key aspect, the transient nonthermal excitation of the spin system, cannot be modeled with a 3TM. Therefore, we expect worse quantitative agreement of a 3TM with the ultrafast dynamics of all subsystems. 

Besides the ASD simulations presented here and the 3TM, another model describing the coupled energy flow dynamics of electrons, spins, and the lattice is the spin-lattice-electron dynamics (SLED) model by Ma {\sl et al.}~\cite{2012Ma}. In contrast with our simulations, the SLED model employs the effective spin temperature~\cite{2010Ma} to calculate the energy flow dynamics. In order to investigate differences between the two descriptions, a comparison analogous to Fig.~\ref{fig6:nonthermal} can be made for the effective spin temperature. In our simulations, we find that on short timescales the energy content calculated using the effective spin temperature can differ from the energy of the spin system calculated using the spin Hamiltonian (see Eq.~\ref{eq:spin_energy}), in particular for iron and nickel and for high fluences. Such deviations from equilibrium relationships indicate a nonthermal behavior of the spin system, as discussed earlier. An additional difference between our model and the SLED model is that the latter includes direct spin-lattice coupling. For these two reasons, we expect qualitatively similar results for the energy flow dynamics but quantitative differences between these two models. 

\section{Summary and conclusions}
\label{sec:summary}
In this work, we investigated the ultrafast lattice dynamics of ferromagnetic Co and Fe using femtosecond electron diffraction. To model the intrinsic energy flow between electronic, magnetic, and lattice degrees of freedom, we combined spin-resolved DFT calculations of the electron-phonon coupling with energy-conserving ASD simulations. We found that for both Co and Fe, the ultrafast spin dynamics have a profound impact on the lattice dynamics, slowing down the lattice heating due to energy transfer into and out of magnetic degrees of freedom. These findings generalize our previous results for Ni~\cite{our_nickel}, highlighting the prominent role of the spin system in the energy flow dynamics of all three elemental 3{\sl d} ferromagnets.

For a full description of the laser-induced dynamics, it is thus essential to take energy flow into and out of the spin system into account. This is achieved with energy-conserving ASD simulations, which simulate the spin dynamics while also accounting for the intrinsic energy flow between electrons, spins, and the lattice. For low and moderate fluences, the ASD simulations yielded excellent agreement with the measured lattice dynamics, as well as a good description of the magnetization dynamics for both Co and Fe. They are thus an important step toward a model for ultrafast demagnetization that is consistent with the responses of electronic, magnetic, and lattice degrees of freedom.

Furthermore, we found that the ASD simulations predict a nonthermal spin system for both Co and Fe on short timescales after laser excitation. For high fluences, the nonthermal state of the spin system can last for several picoseconds, suggesting that particularly for strong excitations, a thermal description of the laser-induced spin dynamics is not sufficient.

Our findings are also of relevance for other demagnetization models since they enable the comparison with the experimental lattice dynamics for all three elemental 3{\sl d} ferromagnets and highlight the importance of a consistent description of the energy flow dynamics. The combination of experiment and theory presented in this work can also be applied to gain insight into the ultrafast energy flow dynamics in other technologically relevant magnetic materials, e.g. magnetic oxides and layered van der Waals materials. In addition, the incorporation of the energy exchange of the spin system in the ASD simulations may prove to be invaluable for the explanation of the behavior of more complex materials and heterostructures in the future. \\
\\

\FloatBarrier

The femtosecond electron diffraction data and the ASD simulation results presented here are available on a data repository~\cite{FeCo_data,Fe_Co_ASD_data}. The code to calculate the electron-phonon coupling parameter and the heat capacities from DFT results is also available, along with results for iron, cobalt, and nickel~\cite{TTM_inputs_code}.

\section*{Acknowledgments}
We thank Victoria C. A. Taylor for assistance during measurements and Reza Rouzegar for sample characterization measurements.
This work was funded by the Deutsche Forschungsgemeinschaft (DFG) through SFB/TRR 227  ``Ultrafast Spin Dynamics" (Projects B07, A08, A09, and A02) and through the Emmy Noether program under Grant No. RE 3977/1, by the European Research Council (ERC) under the European Union’s Horizon 2020 research and innovation program (Grant Agreement No. ERC-2015-CoG-682843), and by the Max Planck Society. H.S.~acknowledges support by the Swiss National Science Foundation under Grant No.~P2SKP2\textunderscore184100. 

\section*{Appendix A: DFT calculations}

The calculations of the electron-phonon energy transfer rates were performed using the DFT code ABINIT \cite{Gonze1997,Gonze1997a,Gonze2009,Gonze2016,Bottin2008}. The
optimized norm-conserving Vanderbilt pseudopotentials were generated using the method of Ref.~\onlinecite{2013Hamann}
and are of generalized-gradient approximation Perdew-Burke-Ernzerhof type~\cite{Perdew:1996}. Sixteen electrons were treated explicitly for Fe, and 17 electrons were explicitly taken into account for Co. The plane-wave expansion of the electronic wave function had a cutoff of 40 Ha for Fe and 50 Ha for Co; 22 electronic bands were calculated for Co and 15 for Fe. These bands were calculated with Fermi occupation featuring a smearing of 0.001 Ha. An unshifted k-point grid of $32\times32\times32$ points was used for both elements. The lattice constant for body-centered cubic (bcc) Fe was set to $\unit[2.756]{\text{Å}}$, which was obtained by relaxing the structure. For hcp Co, we used the experimental lattice constants
$\unit[a=2.5071]{\text{Å}}$ and $\unit[c=4.0695]{\text{Å}}$ \cite{Co_lattice_constant}. To obtain the electron-phonon coupling $G_\mathrm{ep}$, the spin-resolved
electron-phonon matrix elements were computed
as described in Ref.~\onlinecite{2013Verstraete} for a $8\times8\times8$ grid of $q$ points. From the results, we extracted the Eliashberg functions for majority and minority electrons. The electron-phonon couplings and the electronic heat capacities were then calculated as described in Ref.~\onlinecite{2016Waldecker}. Following Ref.~\onlinecite{our_nickel}, we take chemical potential shifts into account and assume particle conservation within each spin type. Band shifts according to the Stoner model are not considered, since our description of magnetization dynamics with ASD simulations is based on the Heisenberg model.

\begin{figure}[bth]
\begin{center}
\includegraphics[width=\columnwidth]{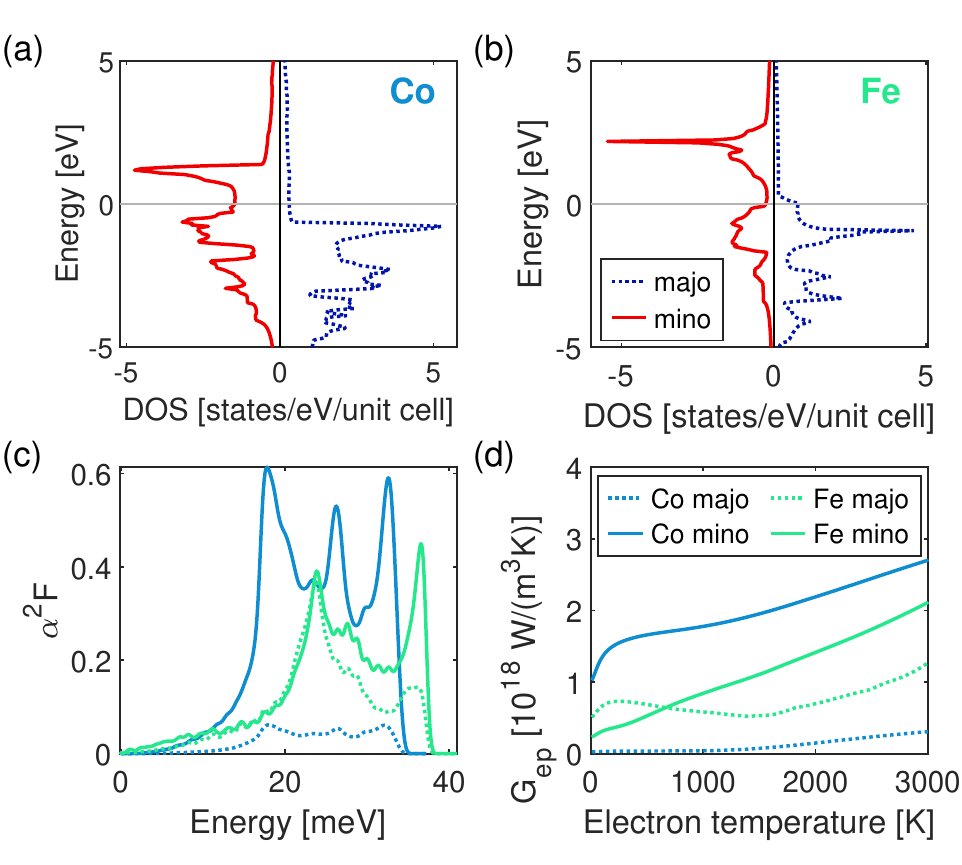}
\caption{Results of the spin-resolved DFT calculations. (a) Spin-split electron DOS of Co. The position of the Fermi level is shown as a gray line. The majority (majo) DOS is shown in dashed blue and the minority (mino) DOS is shown in red. Note that hcp Co has two atoms per primitive unit cell. (b) Spin-split electron DOS of Fe. (c) Majority and minority Eliashberg functions $\alpha^2$F for Co (blue) and Fe (green). The dashed curves correspond to the majority Eliashberg functions and the solid curves represent the minority Eliashberg functions. (d) Majority and minority electron-phonon coupling parameter $G_\mathrm{ep}$ for Co and Fe.}
\label{fig_s1:DFT}
\end{center}
\end{figure}

The results for the electron DOS, the Eliashberg functions, and the electron-phonon couplings are presented in Fig.~\ref{fig_s1:DFT}.
The magnetic moments calculated from the spin-resolved electron DOS, $\unit[1.95]{\mu_\mathrm{B}}$ per atom for Co and $\unit[2.40]{\mu_\mathrm{B}}$ per atom for Fe, are larger than the experimental results of $\unit[1.72]{\mu_\mathrm{B}}$ and $\unit[2.22]{\mu_\mathrm{B}}$ per atom \cite{Kittel}. 

Based on the phonon DOS, we also calculated the MSDs as functions of temperature, as described in Ref.~\onlinecite{Peng}. To increase the accuracy of the calculation, we replaced the phonon DOS in the region below \unit[5]{meV} by a fit with the function $f(x)=ax^2+bx^3$. This ensures that the dominating term for very small phonon wavevectors is quadratic, which corresponds to the correct long-wavelength limit. The results were used to convert transient MSD changes to lattice temperatures [see Figs.~\ref{fig2:exp_results}(a) and \ref{fig2:exp_results}(b)].

\section*{Appendix B: ASD simulations}

\begin{figure}[bth]
\begin{center}
\includegraphics[width=\columnwidth]{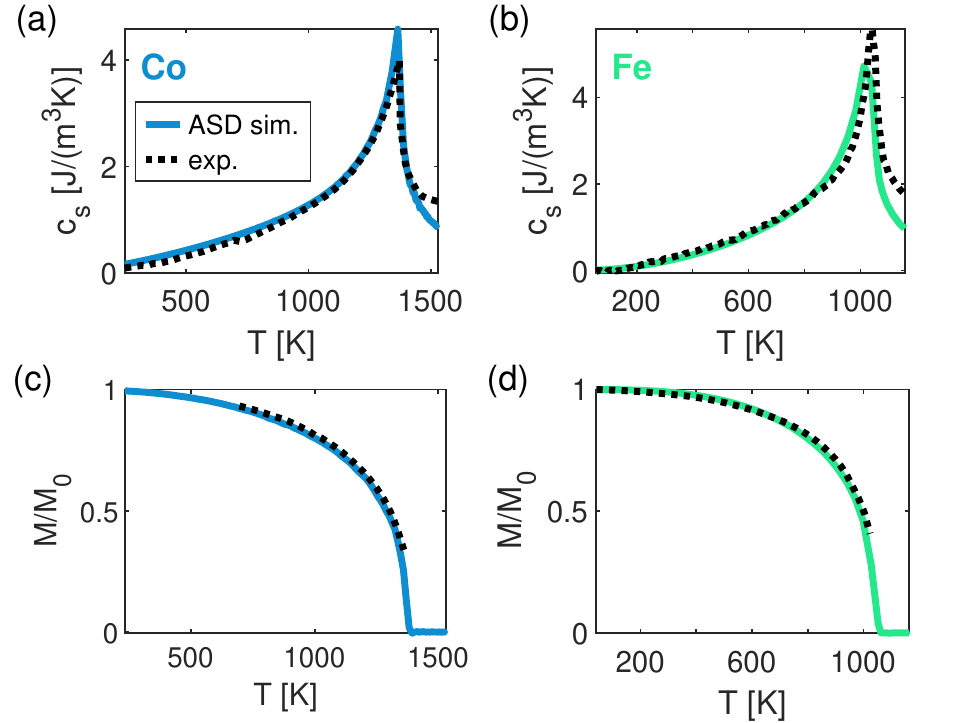}
\caption{ASD simulation results for equilibrium relationships and comparison to literature results. (a) Spin heat capacity of Co. The ASD simulation result (ASD sim.) is shown as a solid blue curve and the experimental result (exp.) is shown as a dashed black curve. The experimental result was obtained based on measurements of the total heat capacity from Ref.~\onlinecite{1998Thurnay}, which were dilation-corrected using the expansion coefficients from Ref.~\onlinecite{CRC_handbook}. To obtain the spin heat capacity, the DFT results for the electronic and lattice contributions were subtracted. (b) Same as (a) but for Fe. (c) Magnetization as a function of temperature for Co. The solid curve shows the ASD simulation result. The dashed black curve is a literature result from Ref.~\onlinecite{1971Crangle}. The magnetization is normalized to its value at \unit[0]{K}. (d) Same as (c) but for Fe.}
\label{fig_s2:ASD}
\end{center}
\end{figure}

\begin{figure*}[t!]
\begin{center}
\includegraphics[width=\linewidth]{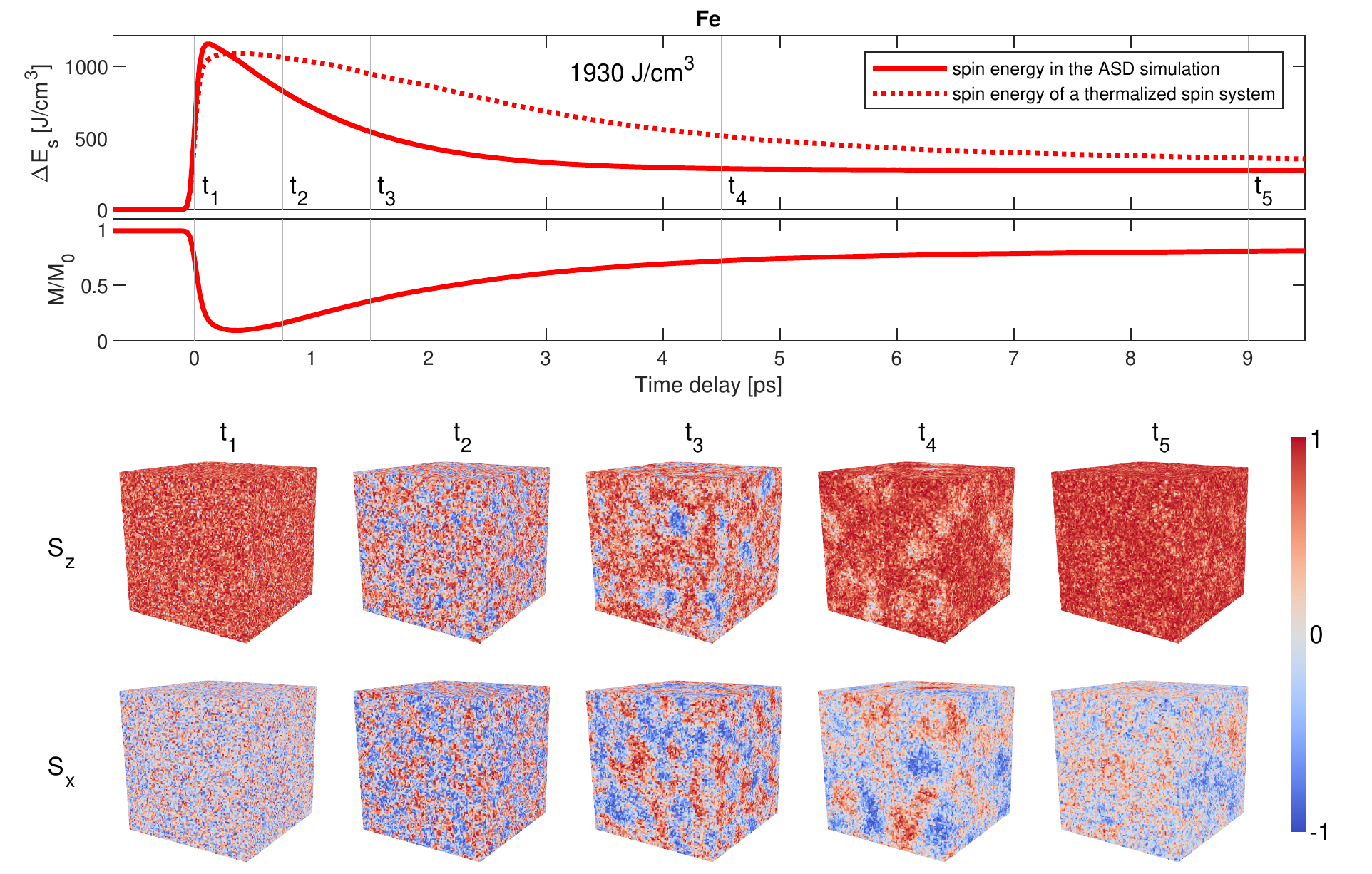}
\caption{Details about the spin configuration at selected time delays from an ASD simulation of Fe at a high excitation density. The upper panel shows the additional spin energy as a function of pump-probe delay (solid curve) for an absorbed energy density of $\unit[1930]{\hspace{1pt}J/cm^3}$. The dashed curve corresponds to the energy content of a thermalized spin system with the magnetization dynamics from the ASD simulations (shown directly below), analogous to Fig.~\ref{fig6:nonthermal}. Note that for illustration purposes, we used a higher damping parameter of $\alpha=0.1$ for the simulations shown here, and a spin system consisting of $100\times100\times100$ spins. Due to the relatively high fluence, the spin system exhibits a nonthermal behavior also on few-picosecond timescales. The lower part of the figure shows the spin configurations at different pump-probe delays. The spin components $S_x$ and $S_z$ are displayed, normalized to 1. $S_y$ behaves analogously to $S_z$ for symmetry reasons. Here only the surface of the cube is visible. The inside of the cube displays an analogous behavior to the surface.}
\label{fig_s3:nonthermal}
\end{center}
\end{figure*}

Atomistic spin dynamics simulations use a classical Heisenberg spin model: 
 \begin{equation}   
\mathcal{H}= - \sum_{i < j} J_{ij} \mathbf{S}_i \cdot \mathbf{S}_j - \sum_{i} d_z S_z^2.
\label{eq:ASD_Ham}
\end{equation}
with $\mathbf{S}_i $ representing a classical, normalized spin vector at site $i$. Each spin couples to its neighboring spin vectors $\mathbf{S}_j$ via the coupling constant $J_{ij}$.
The second term of the Hamiltonian (Eq.~\ref{eq:ASD_Ham}) describes the on-site anisotropy with an easy axis along the $z$ axis and constant anisotropy energy. All parameters are material-dependent and listed in Table~\ref{table:ASD_Parameters}. Except $\alpha$, they are based on Ref.~\onlinecite{Evans_2014}. The simulations are performed on a simple cubic (sc) lattice. This has no significant effect on the energy flow or magnetization dynamics, which was verified directly by comparing simulations of bcc and sc Fe. The reason is that for a given Curie temperature, the different number of neighbors in each case is compensated by a different value of $J$. Note that in contrast with the samples employed in the diffraction experiments, the ground state in the ASD simulations is a single-domain state. Due to the typical time and energy scales of domain wall dynamics, we don't expect a significant influence of the domain structure on the intrinsic energy flow dynamics studied here.
\begin{table}[t!]
\caption{ASD simulation parameters for Co and Fe.}
\label{table:ASD_Parameters}
\begin{tabular}{ l |lll|lll}
\hline 
\hline
& Co&&Units&Fe&&Units\\ \hline
$J$ &$6.324$&$\times 10^{-21}$&[J]&$4.8 $&$\times 10^{-21}$ &[J]\\
$d_z$ &$0.67$&$\times 10^{-23}$&[J]&$0.5$&$\times 10^{-23}$&[J]\\
$\mu_{\rm{s}}$& 1.72 &&$[\mu_{\rm{B}}]$&2.2&&$[\mu_{\rm{B}}]$\\
$\alpha$&$0.01$&&&$0.005$&&\\
\end{tabular}
\end{table}
By solving the s-LLG equation
\begin{equation}
\frac{(1+\alpha^2)\mu_\mathrm{s}}{\gamma}\frac{\partial \mathbf{S}_i}{\partial t} = - \left( \mathbf{S}_i \times \mathbf{H}_i \right) - \alpha  \left( \mathbf{S}_i \times \left(\mathbf{S}_i \times \mathbf{H}_i \right) \right)
\label{eq:llg}
\end{equation}
numerically, the dynamics of the system are calculated \cite{Nowak2007}. Here $\unit[\gamma=1.76 \cdot 10^{11}]{\frac{1}{Ts}}$ refers to the gyromagnetic ratio and $\mathbf{H}_i$ describes the effective field derived via $\mathbf{H}_i= -\frac{\partial \mathcal{H}}{\partial \mathbf{S}_i}$. The material-dependent and phenomenological damping parameter $\alpha$ determines the coupling strength of the spin system to the electron system and thus the energy transfer rate between the two heat baths.
In order to simulate the effects of finite temperatures, a Langevin thermostat is included by adding a field-like stochastic term $ \boldsymbol{\zeta}_i$ to the effective field $\mathbf{H}_i= \boldsymbol{\zeta}_i(t) - \frac{\partial \mathcal{H}}{\partial \mathbf{S}_i}$. The added noise term has white noise properties~\cite{Atxitia2009}:
\begin{equation}
\langle \boldsymbol{\zeta}_i(t) \rangle = 0 \quad \text{and} \quad \langle \boldsymbol{\zeta}_i(0) \boldsymbol{\zeta}_j(t) \rangle = 2 \alpha k_\text{B} T_{\rm{el}} \mu_\mathrm{s} \delta_{ij}\delta(t)/\gamma.
\label{eq:noise-correlator}
\end{equation}
In order to better reproduce experimentally measured equilibrium properties such as magnetization and heat capacity, we make use of a rescaled temperature model, which utilizes a slightly modified electron temperature $T_\text{sim}$ for the noise generation. Figure~\ref{fig_s2:ASD} presents simulation results for equilibrium properties and a comparison to literature values. In addition to the equilibrium properties, the rescaled temperature model also yields a better description of nonequilibrium dynamics~\cite{Evans2015}. Further details are available in Refs.~\onlinecite{Evans2015,our_nickel}.

A major advantage of ASD simulations is that they are not limited to a thermal description of the spin system, since the spins are simulated directly. Figure~\ref{fig_s3:nonthermal} shows the evolution of the spin energy content and a direct visualization of the simulated spin dynamics for a relatively high fluence of $\unit[1930]{J/cm^3}$. In addition, for illustration purposes, a simulation with a higher damping of $\alpha=0.1$ is shown. A higher  damping leads to a larger disorder of the spin system directly after excitation, however, the qualitative behavior displayed in Fig.~\ref{fig_s3:nonthermal} is also observed for lower values of the damping parameter at high fluences.

The comparison of the simulated spin energy content (solid curve) to the energy content of a thermalized spin system with the simulated magnetization dynamics (dashed curve) reveals that the spin system remains in a nonthermal state for several picoseconds. At short time delays, the nonthermal state is characterized by a relatively large spin energy content compared with the demagnetization amplitude. The behavior reverses on longer timescales. Further insights on these nonthermal states can be gained from the visualization of the spin dynamics. The instantaneous spin configuration is illustrated exemplarily for several delays after excitation. During and directly after excitation, e.g., at $t_1=\unit[0]{ps}$, there is significant short-range disorder in the spin system, i.e., significant misalignment between neighboring spins. According to the Heisenberg Hamiltonian, this comes with a significant energy cost and thus leads to the relatively large energy content of the spin system. On longer timescales, the magnetization recovers. However, for high fluences/strong demagnetization, domains form. This is already visible at 
$t_2=\unit[0.75]{ps}$. There are areas with significant magnetization in which the spins point predominantly in the $x$ or $y$ direction (note that in Fig.~\ref{fig_s3:nonthermal}, each spin is normalized such that $S_x^2+S_y^2+S_z^2=1$).
This behavior is similar to spin simulation results reported in Ref.~\onlinecite{Kazantseva2007}. Within a domain, spins are parallel. Therefore, the energy cost of this spin configuration is relatively low. Nevertheless, due to the different directions of the magnetization, the global magnetization is reduced.
In the beginning, the domains are relatively small. As time progresses, the domains become larger (see $t_3=\unit[1.5]{ps}$ and $t_4=\unit[4.5]{ps}$) and eventually disappear (see $t_5=\unit[9]{ps}$) as the spin system approaches thermal equilibrium. Note that for low fluences, domain formation as illustrated in Fig.~\ref{fig_s3:nonthermal} does not occur, since it requires significant initial disordering of the spin system. The initial nonthermal disorder of the spin system, visualized in Fig.~\ref{fig_s3:nonthermal} for $t_1$ and characterized by a large number of high-energy spin excitations, occurs for all fluences (see also Fig.~\ref{fig6:nonthermal}).

\bibliography{main}

\end{document}